\pdfoutput=1
\documentclass[review]{elsarticle}
\usepackage{amsmath,amssymb,amsfonts}
\usepackage{graphicx}
\usepackage{textcomp}
\usepackage{url}
\usepackage{subcaption}
\usepackage{multirow}
\usepackage{placeins}
\usepackage{array}
\usepackage[title]{appendix}
\usepackage{nomencl}
\usepackage{algorithm}
\usepackage{algpseudocode}
\makenomenclature
\usepackage{IEEEtrantools}
\newcolumntype{P}[1]{>{\centering\arraybackslash}p{#1}}
\newcolumntype{C}[1]{>{\centering\arraybackslash}m{#1}}
 
\journal{Journal of \LaTeX\ Templates}

\begin{document}

\begin{frontmatter}

\title{COVID-19 Digital Contact Tracing Applications and Techniques: A Review Post Initial Deployments}



\author[address1]{Muhammad Shahroz}
\ead{mshahroz.msee18seecs@seecs.edu.pk}
\author[address1]{Farooq Ahmad}
\ead{fahmad.msee17seecs@seecs.edu.pk}
\author[address1]{Muhammad Shahzad Younis}
\ead{muhammad.shahzad@seecs.edu.pk}
\author[address2]{Nadeem Ahmad}
\ead{nadeem.ahmed@cybersecuritycrc.org.au}
\author[address3]{Maged N. Kamel Boulos}
\ead{mnkboulos@ieee.org / mnkboulos@mail.sysu.edu.cn}
\author[address4]{Ricardo Vinuesa}
\ead{rvinuesa@mech.kth.se}
\author[address5]{Junaid Qadir}
\ead{junaid.qadir@itu.edu.pk}


\address[address1]{School of Electrical Engineering and Computer Science, National University of Sciences and Technology, Islamabad, 44000, Pakistan.}
\address[address2]{Cyber Security Cooperative Research Centre, (CSCRC), University of New South Wales Sydney, New South Wales, Australia.}
\address[address3]{School of Information Management, Sun Yat-sen University, East Campus, Guangzhou, 510006, Guangdong, China.}
\address[address4]{Department of Mechanical Engineering, KTH Royal Institute of Technology, Stockholm, Sweden.}
\address[address5]{Department of Electrical Engineering, Information Technology University, Lahore, Pakistan.}

\begin{abstract}
The coronavirus disease 2019 (COVID-19) is a severe global pandemic that has claimed millions of lives and continues to overwhelm public health systems in many countries. The spread of COVID-19 pandemic has negatively impacted the human mobility patterns such as daily transportation-related behavior of the public. There is a requirement to understand the disease spread patterns and its routes among neighboring individuals for the timely implementation of corrective measures at the required placement.  To increase the effectiveness of contact tracing, countries across the globe are leveraging advancements in mobile technology and Internet of Things (IoT) to aid traditional manual contact tracing to track individuals who have come in close contact with identified COVID-19 patients. Even as the first administration of vaccines begins in 2021, the COVID-19 management strategy will continue to be multi-pronged for the foreseeable future with digital contact tracing being a vital component of the response along with the use of preventive measures such as social distancing and the use of face masks. After some months of deployment of digital contact tracing technology, deeper insights into the merits of various approaches and the usability, privacy, and ethical trade-offs involved are emerging. In this paper, we provide a comprehensive analysis of digital contact tracing solutions in terms of their methodologies and technologies in the light of the new data emerging about international experiences of deployments of digital contact tracing technology. We also provide a discussion on open challenges such as scalability, privacy, adaptability and highlight promising directions for future work.
\end{abstract}


\end{frontmatter}


\section{Introduction} \label{sec:Introduction}
The Severe Acute Respiratory Syndrome Coronavirus 2 (SARS-CoV-2) virus and the associated disease designated as coronavirus disease 2019 (COVID-19), first reported in China in late 2019, represents the most significant public health threat in the last 100 years. This disease has spread like wildfire across the globe. The pandemic response has been mixed. Some countries responded proactively and effectively, while others botching their responses. One way to limit the spread of the virus is to ensure strict lockdowns---however, this comes at a significant economic cost due to industries being shut down and workers losing jobs, and there is a risk that the spread will start anew once the lockdowns are lifted \cite{imperial}. 

During the early days of the pandemic, human mobility was greatly affected. People started taking self-isolation measures in their homes. However, it resulted in the shutting of the transportation industry. Most countries have focused on developing smart lockdown strategies using various technological solutions to combat COVID-19 \cite{latif2020leveraging,chamola2020comprehensive}. Overall timely preventive measures and a mix of high and low technological solutions made it possible for some countries (such as China, South Korea) to minimize the disease's spread. These countries relied on putting international travelers under surveillance and quarantine. Even low-risk individuals were asked to observe quarantine, and their physical location was either observed by the government operators or cell tower location data was used to monitor the quarantine conditions. Identified cases were kept in special COVID wards for about a fortnight, depending on health infrastructure availability.

The use of face masks among the masses can drastically reduce the virus's spread among healthy individuals \cite{facemasks}. However, the chances of the spread can still be high. Contact tracing is a technique to identify individuals who have possibly come in close contact with an infected person while that person was the carrier of the viral pathogens. Contact tracing is a time-tested technique employed successfully to control and monitor historical outbreaks of diseases like HIV, Ebola, and measles. The traditional way of contact tracing is the \textit{manual contact tracing technique}, used to identify the close meetups of the infected person \cite{2003contact}. However, the manual contact tracing technique has two significant limitations: (1) it requires a sizeable trained workforce to conduct these manual interviews; and (2) it cannot identify individuals that are not known to the infected person but have come in close contact (e.g., while using public transport or dining in restaurants). Furthermore, manual contact tracing is a hectic process that requires a centralized, coordinated effort to identify at-risk close contacts of a COVID positive individual.  

Due to the various limitations of manual contact tracing, technologists, in consultations with the epidemiologists, are now overwhelmingly supplementing classical contact tracing techniques with \textit{digital contact tracing techniques} \cite{Anglemyer2020}. Digital contact tracing techniques typically depend on apps installed on smart mobile phones \cite{mobileapps}. These contact tracing apps trace individuals' meetups by either using a local Bluetooth connection or the global network of the Global Positioning System (GPS) for location tracking. 
In the current crisis, many IoT-based contact-tracing apps have been deployed \cite{2003contact,appsurvey} or are in the development phase under the collaborations of government and tech industries \cite{app1,app2,app3}. 
Countries like Singapore, South Korea, Israel, Italy, Germany, and China have fully implemented digital contact tracing. Other countries do not have substantial adoption due to policy issues and concerns about consumer privacy and legal rights. Big tech companies like Apple and Google \cite{app3} have also joined hands to accelerate the effort on expanding the capabilities of existing tracing frameworks. Digital contact tracing platforms are thus emerging as an essential component of global response against the COVID-19.  


However, such an approach is not a panacea or a silver bullet. The current generation of digital contact tracing apps is facing several issues that limit their effectiveness. This includes low app adoption rates, low mobile phone penetration, privacy and trust issues, potentially high false-negative rates (i.e., the app fails to register a close contact with an infected individual \cite{introlink}), and the reliance on the tracing apps on Bluetooth for proximity calculations. In this paper, we present a comprehensive analysis of digital contact tracing and its supporting IoT framework. Furthermore, we elaborate on the challenges associated with the digital contact tracing solutions in terms of their methodologies and technologies in the light of the new data emerging about international experiences of deployments of digital contact tracing technology. We also discuss open challenges such as scalability, privacy, adaptability and highlight promising directions for future work. 

The rest of the paper is structured as follows. The background of digital contact tracing, which includes discussion on multiple types of contact-tracing architecture, review of different existing communication technologies in contact tracing, and the IoT framework of contact tracing, is presented in Section \ref{back}. The challenges associated with the digital contact tracing application are discussed in Section \ref{Challenges}. The international efforts in digital contact tracing are discussed in Section \ref{IntApp}. The current research gaps in the literature are identified in Section \ref{gaps}. The emerging communication technologies that can be used in digital contact tracing are discussed in Section \ref{Emerg_Comm}, followed by concluding remarks in Section \ref{Conclusion}.

\section{Background} \label{back}

Digital contact tracing is a framework in which smartphones register close contacts with other smartphones, running the same contact tracing app. In this section, the architecture of the digital contact tracing framework is discussed. Furthermore, the communication technologies which are currently used for digital contact tracing are discussed. Moreover, digital contact tracing solutions using IoT are also discussed in this section.   

\subsection{Centralized vs. Decentralized Digital Contact Tracing Architecture} \label{architecture}

\begin{figure*}[htbp]
\centering
\includegraphics[width=\textwidth]{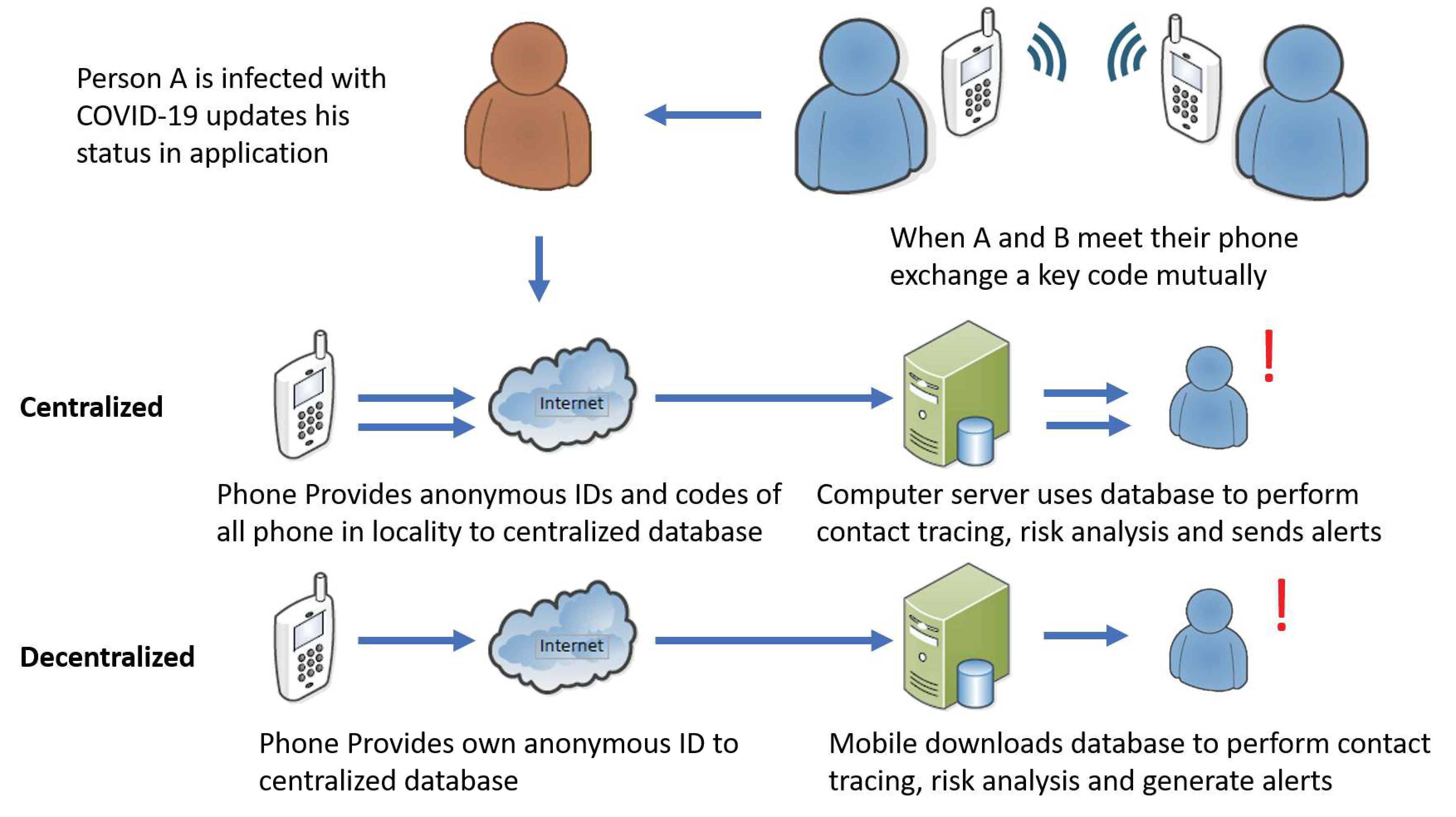}
\caption{Two types of contact-tracing architecture}
\label{fig:cent}
\end{figure*}

Broadly speaking, contact-tracing applications are of two types. In \textit{centralized contact-tracing}, mobile phones share their anonymous IDs to a central server maintaining a centralized database, and the server uses this database to perform contact-tracing, risk-analysis and alerts notifications to the users. In \textit{decentralized contact-tracing}, on the other hand, mobile phones, instead of a centralized server, perform contact matching and notification by downloading the contact database from the server. 
A graphical illustration of these contact-tracing architectures is presented in Figure \ref{fig:cent}. In centralized contact-tracing, the detection is performed in a centralized server. In contrast, in decentralized contact-tracing, each user smartphone acts as a local server that shares only the infected individuals' data to the centralized server, and then the smartphones will fetch this data periodically from the server and do contact matching locally. An example of this decentralized contact-tracing architecture is Apple-Google platform \cite{Google_Apple}. Note that in decentralized contact tracing, only the data of an infected person are shared with the centralized server, and contact matching is performed locally---a fact that makes the user privacy more effective as compared to the centralized contact tracing in which data of all individuals are shared with the centralized server. However, there is still a vigorous debate on the privacy and security aspects and use of these contact-tracing architectures.

\subsection{Communication Technologies Used in Digital Contact Tracing} \label{tech}

\begin{figure*}[htbp]
\centering
\includegraphics[width=\textwidth]{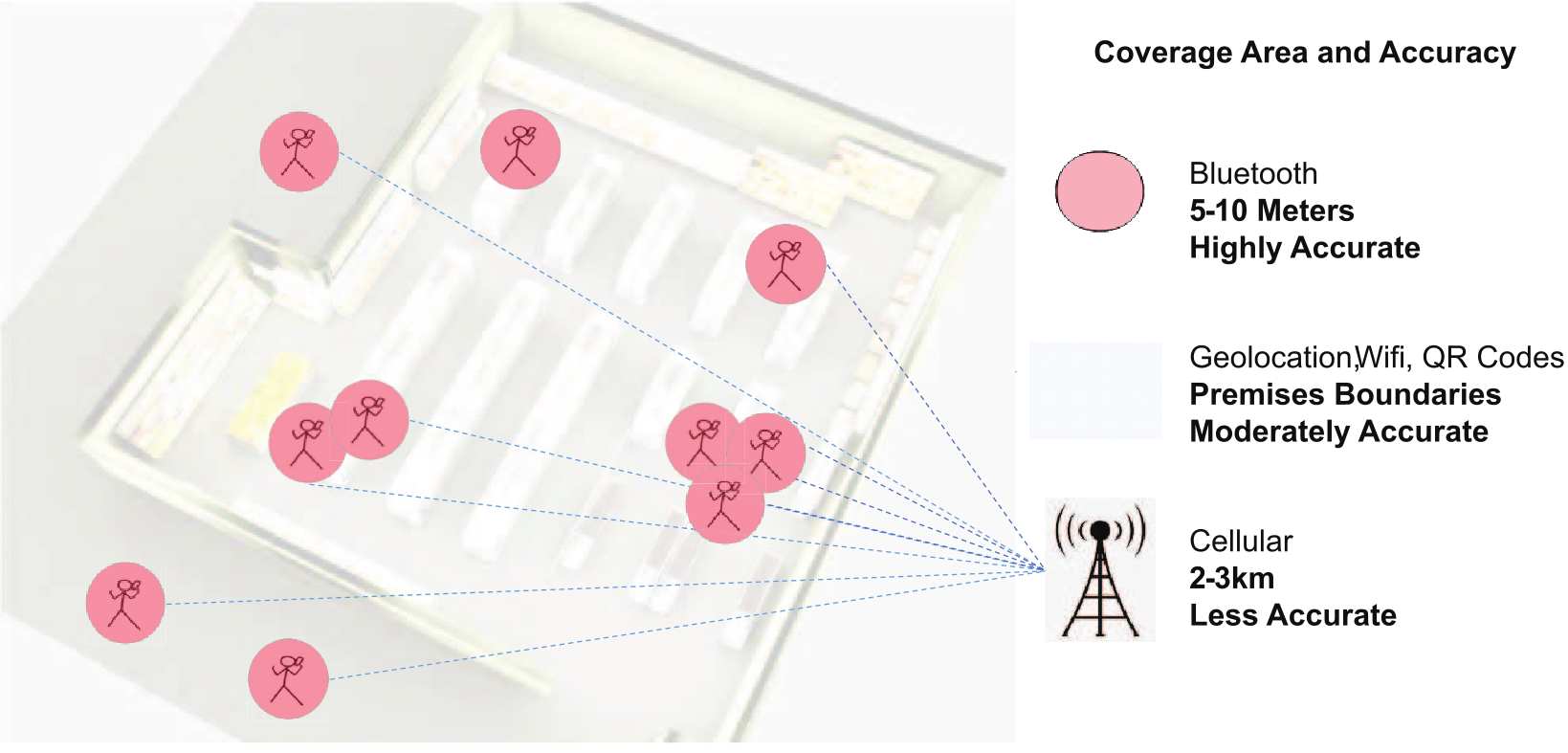}
\caption{Coverage Area and Accuracy Comparison for Different Sensing Infrastructures }
\label{fig:CT}
\end{figure*}

To enable technlogy driven social distancing, many communication technologies are adopted to enhance the effectiveness of the digital contact tracing solutions \cite{IEEE_Access_COVID_Survey_1}, e.g., WiFi, Bluetooth, GPS, QR Codes, and Zigbee. A brief explanation of these communication technologies used in contact-tracing applications is discussed below, and their coverage ranges and accuracy is shown in Figure \ref{fig:CT}.  

\begin{enumerate}

\item{WiFi}

WiFi is a communication technology which is very effective for contact-tracing purposes especially in an indoor environment such as multi-story building, airports, alleys, and parking garages \cite{WiFi}. WiFi provides high accuracy in an indoor environment compared to GPS or other satellite-based technology, useful in an indoor environment. In a WiFi system, a wireless transmitter known as WiFi Access Point (WAP) is required to communicate with its user devices. One of the potential WiFi technology applications used in contact tracing is positioning \cite{WiFiPos1,WiFiPos2}. This is very useful in an indoor environment when a crowd is buildup in pandemic times, especially in railway platforms and airports. It is also convenient to use the WiFi hardware facilities due to their low-cost maintenance and easy deployment.     

\item{Bluetooth}

Bluetooth is a wireless communication protocol that is present in almost every modern mobile phone. There are several versions of Bluetooth protocols. Bluetooth Low Energy (BLE) is very popular in applications due to its less energy consumption and low cost. As the contact-tracing apps are required to run continuously for logging the contacts, BLE low battery consumption is very well suited. However, BLE has a short-range coverage, mostly indoors. One of the main advantages of Bluetooth technology is that it can connect a device to multiple devices without requiring any access point and forming an ad-hoc called piconet \cite{piconet}.  

In Singapore, the government has developed an open-source contact-tracing protocol called BlueTrace, which is employed in the TraceTogether app \cite{app4}. The methodology is very simple; every time the contact-tracing app comes within the proximity of another app, it will locally save their information mutually. They may volunteer to share this contact information later with the health department. Most of the contact-tracing apps utilize Bluetooth for contact-tracing.

\item{Global Positioning System}

The GPS navigation system uses a network of satellites to locate the exact position of the GPS-enabled devices. Modern smartphones are GPS enabled, which can be used for contact tracing as well. The other benefit of GPS is its global availability. Many countries like Israel\footnote{\url{https://govextra.gov.il/ministry-of-health/hamagen-app/download-en/}.} and Norway\footnote{\url{https://helsenorge.no/coronavirus/smittestopp last accessed 2020/05/25}.} are using GPS-based contact-tracing mechanism.

GPS technology can also be used to limit the physical contact between people, e.g., customers can shop online and get the product delivered to their homes using Unmanned Aerial Vehicles (UAVs) based on GPS technology. Many big retail and logistic corporations are investing in  UAVs to deliver the products to customer homes like Amazon and DHL. Therefore, the social distancing between the public can be significantly enhanced using GPS technology. 

\item{QR Codes}

Another method of contact tracing is the usage of QR codes where a user will be manually contributing to the database by taking a picture of visual computer-aided code at multiple places of business. Mobile phone application automatically reads hidden geolocation of these QR codes and populates the database with user details. If a person is tested positive, their contacts at places they have visited can be identified using contact-tracing based on QR codes. Such techniques are globally used, with China being a prime example that has a far higher rate of adoption of QR codes \cite{chineseQR}.

\item{Zigbee}

Zigbee is also a potential technology that can be used in maintaining social distancing. Zigbee is a standard-based wireless communication technology used for low-cost and low-power wireless networks \cite{zigbee}. Zigbee-based devices can communicate with each other in the range of about 65 feet (20 meters) and can take unlimited hops. The Zigbee control hub can determine the user's location, which can be used for crowd control. Therefore, the Zigbee communication technology can be used for contact-tracing purposes to avoid the spread of the virus.       

\end{enumerate}

\subsection{Digital Contact Tracing with IoT} \label{IoT}

Internet of Things (IoT) is a system of computing devices that has the ability to transfer data over a network without requiring human-to-human or human-to-computer interaction. In the current epidemic, the number of infected cases is increasing day by day; therefore, it is becoming increasingly hard to rely only on manual contact tracing. 

Digital contact tracing based on IoT can enable a scalable, automated contact-tracing system that can cope with the ever-increasing workload of contact tracing. Digital contact tracking is a technique used to aid the contact-tracing process, which detects the contacts when a person is in the proximity of an infected person. This can be done either by using GPS location or by Bluetooth signal. The risk of exposure to the infection depends upon how close a person comes close to the infected person (say less than 1 meter) and the duration of this contact. However, proximity tracking does not provide a complete assessment in tracking contacts with an infected person because of obstacles between two persons, such as a wall or being enclosed in personal protective equipment. 

Multiple types of contact-tracing frameworks have been established following regional compatibility, acceptability, privacy, and security laws. Most of these solutions depend on sensors to either identify the user's close contacts to understand the spread of the epidemic or track the location of the users (tracking solutions). These applications require different permissions or sensory input to either flag or alert users of imminent virus transmission threats. Tracking platforms like Google COVID-19 Community Mobility Reports\footnote{\url{https://www.google.com/covid19/mobility/}.} and Apple COVID-19 Mobility Trends Reports\footnote{\url{https://www.apple.com/covid19/mobility}.}, take up data from millions of users with their consent and aggregate this information anonymously. This is done to track people and crowds' movement, to know about the supposed hotspots, and, most of all, to track the effectiveness of lockdowns in different places and countries. These contact-tracing apps are meant for healthy individuals who are keen to know about their susceptibility to getting infected if they have come across an individual later diagnosed with COVID-19. With the application of these digital contact tracing apps, the passenger flow can be control in the passenger waiting areas by forecasting the transportation data \cite{ChinaTrans}. 

More aggressive solutions have also been developed to geofence the individuals in quarantine using dedicated smartphone applications or more straightforward interaction with the user using a phone call or text message exposing the geolocation of the user. An IoT hardware-based solution is presented in \cite{IoTSol}, in which the information on movement and contacts of objects are captured using RFID tags. Such implementation is more obligatory than voluntary, with noncompliance deemed as an offense. Recently a platform named as BubbleBox is presented in \cite{bubbleBox}, in which the system of integrated IoT devices and a software platform is used to limit and detect further outbreaks of COVID-19 infections. A BubbleBox device, a wristband, traces contacts under the safe social distance. With a web-app, the users can pair their identity with their device, as well as report their symptoms. In this way, they offer the authorized medical personnel a quick way to understand the infection's spread, monitor who needs to be tested and quickly contact patients. Finally, the collected data, anonymized, can help researchers understand trends about the spreading of the infection. 

Usability is also crucial to increase the number of people who will use the system and, thus, maximize the coverage of the contact tracing. Finally, as different apps and systems to aid in contact tracing become popular, such systems will be effective only if they are interoperable. Therefore, there is a need to establish a standard for the collected data.

\begin{figure}[htbp]
\centering
\includegraphics[keepaspectratio,width=9cm]{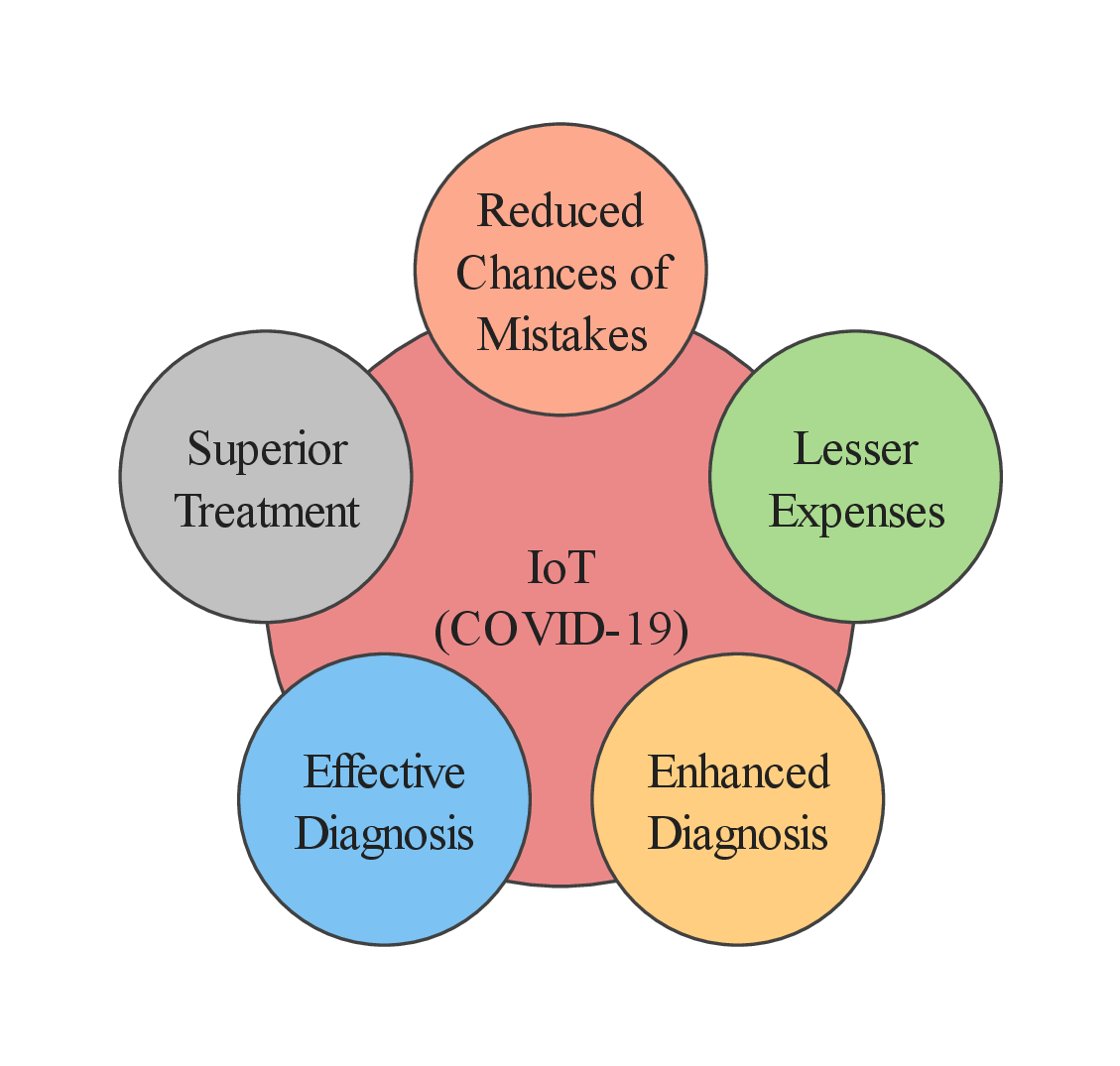}
\caption{Key Merits of using IoT solution in combating COVID-19}
\label{fig:IoTMerits}
\end{figure}

IoT is an innovative technology that can help in tracking all high-risk patients during the quarantine period. Figure \ref{fig:IoTMerits} shows the critical merits of IoT for COVID-19 pandemic. With the successful implementation of this technology, we expect an improvement in medical and contact tracing staff's efficiency with a reduction in their workload. 

\section{Challenges} \label{Challenges}

Since the COVID-19 crisis emerged, everyone is worried about when this pandemic will stop, and the worldwide lockdown, which caused an unprecedented economic crisis, will come to an end. Although the use of digital contact tracing apps has significantly increased, these contact-tracing apps are still facing issues and challenges apart from the benefits they have delivered. In this section, we will discuss some of the challenges that impede the effectiveness of these apps.       

\subsection{User Adaptability}

The success of digital contact tracing largely depends on user adaptability. These apps face a low penetration rate among the masses \cite{covidTracker}, e.g. Australia's COVIDSafe has a 28.6\% penetration rate, Singapore's TraceTogether has a 25\% penetration, India's Aarogya Setu has 12.05\%, Turkey's Hayat Eve Sığar has 17\% and UK's NHS COVID-19 App has 28.5\%, which are among the highest and many are below the 10\% adoption e.g. Japan's COCOA has 6.09\% and France's TousAntiCovid has 3.58\%. 
For a penetration rate of say 25\%, when two persons meet at random, there is only a 6.25\% (0.25 $\times$ 0.25) chance that both persons have installed the app on their mobile. This is the best possible scenario in which both persons have an app running on their mobile devices. It is a challenge to achieve high user adaptability, given that the users are skeptical about the privacy issues surrounding the use of data generated through these apps.  

\subsection{Low Smart Phone Penetration}
Another reason for less adaptability of these contract tracing apps is the low penetration of smartphones \cite{smartphone}. As most of the digital contact tracing apps are based on smartphones, users with older devices (or smartphones running out-of-support versions of mobile operating systems) are unable to use these apps. The smartphone penetration varies with countries, e.g., 24\% in India, 81\% in the USA, and 95\% in S. Korea \cite{countries}. Even in a country like the United States of America, 2 out of 10 individuals do not even use smartphones; many of those who own such a device may opt not to install the tracing app on their device. The digital contact tracing is thus only targeting a fraction of the population that has access to supported devices. 

\subsection{Global Reach}
With the presence of automated and manual contact tracing frameworks, it is critical to gauge the level of effectiveness of both systems. Manual contact tracing requires reliable government infrastructure and higher funds to run the whole operation for a long duration. However, automated contact tracing requires a modern and high-tech infrastructure. The better the technology infrastructure of a country, the better the digital contact tracing will be. We have already discussed that smartphone penetration and user adaptability are risks affecting the success of these apps. Moreover, the elderly and people from low socioeconomic communities are more vulnerable to being affected by the outspread. However, these same groups are the one that is most unlikely to own a smart device. From a global perspective, digital contact tracing may be suitable for developed countries. However, in developing and underdeveloped countries, digital contact tracing frameworks may not achieve their full potential \cite{DataCovid19}.

\subsection{Privacy and Ethical Issues} \label{PrivacyIssues}

The contact-tracing process includes gathering privacy-sensitive information of individuals. However, privacy-conscious people generally show reluctance to share their sensitive information, obstructing the whole contact-tracing procedure. Therefore, privacy-preserving contact-tracing apps are more likely to be adopted by a large user base.
In the literature, privacy-preserving contact tracing has been proposed even before the COVID-19 pandemic. Proposals include Efficient Privacy-preserving contact-tracing for Infection Detection (EPIC) \cite{EPIC} and ENcounter-based Architecture for Contact Tracing (ENACT) \cite{ENACT}. These techniques were developed for privacy-preserving contact tracing and can be useful in the current scenario as well. 
Most of the well-known digital contact tracing apps have some level of privacy built-in. The individuals' smartphone identities are made anonymous before sharing with other users and to the centralized database. The database can only use this anonymous identity for contact matching, risk analysis, etc.

Management of this health crisis of unprecedented magnitude requires desperate measures, but it cannot compromise civil and privacy rights \cite{Vinuesa2020privacy,Dignum2019privacy}. This may result in a restricted contact-tracing framework with lower efficiency. Such problems with data privacy affect the contact-tracing application of a single state or a country and questions the inter-operability of multiple contact-tracing frameworks of different countries with a far higher number of people crossing borders. Under these circumstances, organizations like the European Commission have called upon a common approach among all European governments. A standard policy has been devised which provides multiple recommendations, i.e., all citizen level data will be encrypted and will be erased once the pandemic is near its end, use of contact-tracing app should be voluntary instead of compulsory. Most of all, the application will avoid using location-based data tracking instead of proximity sensing will be used among users.

All these interventions, which, in a way or another, depending on the existence of a digitized system, always pose as a threat to ethics, privacy, and equality. Such subsystems always remain unreachable to some masses which do not have that level of digital literacy irrespective of the user level of digital readiness. For example, everyone should be able to access these technology-based solutions irrespective of what mobile phone they own or which version of the mobile operating system their devices can support \cite{ethics}.

\subsection{Technology Limitations and Transparency} \label{Technology}
Most of the digital contact tracing apps employ the received signal strength of Bluetooth messages for proximity estimation.  This is not a very reliable mechanism for distance estimation and may result in erroneous measurements. An error in distance estimation, in turn, can simulate either panic or a false sense of safety in users. Bluetooth-based proximity sensing with its faults can alert of an infection risk even if contacts are separated by a wall, or it can induce a false level of security even in the proximity of an infected individual who is either not using the application at the moment or if the proximity-based sensing fails to register the contact. 

There are many instances of governments being questioned for the transparency of their solutions. A significant amount of effort has been spent on the development of such systems. Still, they are liable to rejection by individuals if they are not satisfied with the level of privacy and security. This can affect the trust they have in local government and the healthcare system to uphold their ethical obligations.

\section{International Contact Tracing Approaches} \label{IntApp}

The COVID-19 pandemic has taken the world by surprise, mainly due to its spread rate. Governments are taking every possible measure to protect their citizens and their economies from this disease's effects. Many Countries have done different modeling studies, explicitly evaluating the COVID-19 pandemic in terms of mathematical modeling of disease transmission rate and spread pattern \cite{Ferretti2020,Fournet2016,Kucharski2020}. 

Most of the countries have adopted digital contact solutions to support the manual tracing processes. These apps vary in the way contact tracing is performed. Some of the apps use GPS to track the users' movements, while others use a more privacy-preserving design based on Bluetooth advertisements to register close contacts. Some governments have mandated the use of their apps, while others encourage voluntary adoption. To build trust, many apps have their source code released for public scrutiny. The list of apps used by 42 different countries with their technology adoption \cite{mobileapps} is presented in Table \ref{tab:countrieslist}.\footnote{See also \url{http://healthcybermap.org/WHO_COVID19/#8}.}
Most of these international digital contact tracing apps are based on Bluetooth and GPS technology, while some also use QR.

There is a more significant concern about using such tracing applications because of the possibility of fine-grained location tracking and access to private health records. It is essential for an application to be acceptable to the masses that its working and policies are transparent. While some apps (mostly from the European countries including Austria, Finland, Denmark, Norway, Estonia, and United Kingdom) are transparent in terms of their open-source design and codes, many contact-tracing applications such as those used by governments of Algeria, Kuwait, and Tunisia have been questioned by Amnesty International for their lack of transparency. The contact-tracing application of Qatar is mandatory for citizens and alarmingly requires access to mobile phone photos. The application of UAE is decentralized on the surface, yet authorities can fine the individuals avoiding usage or registration of this application. Furthermore, the contact-tracing application of Iran has been taken off from Google Play Store for gathering unnecessary data. However, recent analysis has shown that the effectiveness of these digital solutions can only be increased once the population participates in the contact-tracing process effectively \cite{kim2020contact}. We can learn from Australia and Singapore's experience about why their contact-tracing apps are unsuccessful at large and what steps can be taken to make these successful apps \cite{Aus_Sing}. Google and Apple also announced their partnership in the "spirit of collaboration", and build a joint effort to enable the use of Bluetooth technology to help governments and health agencies to reduce the spread of the virus \cite{Google_Apple}. 

In Ref. \cite{vinuesa2020socio}, Vinuesa et al. have proposed a socio-technical framework to evaluate the suitability of digital contact tracing applications. The authors of this work belong to the Scandinavian countries, Sweden and Denmark. More specifically, the framework proposed by Vinuesa et al. comprises 19 criteria that judge the impact of digital contact tracing apps that can be grouped into three categories (impact on citizens, technology, and governance). The authors propose that the technology should be centralized and privacy-preserving (an example architecture is decentralized privacy-preserving proximity tracing (DP-3T)). The technology should use local and temporary encrypted storage. The app should be easy to deactivate or remove and should have an open-source code. In terms of governance, the app should be preferably owned by the state or a health agency rather than from a private/commercial entity. Open data governance is preferable to opaque settings. The authors recommend that the app should be voluntary and that the app should not be made mandatory to attend certain places. A \emph{sunset clause}—which specifies an end date when the collected data will be destroyed unless extended by explicit processes—is preferable. The framework also recommends concurring with the European Data Protection Board (EDBP) guidelines that users should have a right to contest decisions or demand human intervention. Based on this framework, the authors analyze existing applications and describe how applications such as Stopp Corona (app from Austria), NHS COVID-19 (initial version proposed by the UK), and TraceTogether (app from Singapore) all have low scores in governance). The authors show that the Austrian app has maximum compliance with their suggested guidelines while NHS COVID-19 had the least compliance of the three considered apps. Therefore, there is a need for a tool that can measure the effectiveness of these apps in controlling the spread of the pandemic. Recently a SIR (Susceptible-Infectious-Recovered) model is presented that can test the impact of contact tracing apps in different scenarios using demographic, and mobility data \cite{Ferrari2021}. 

\FloatBarrier
\begin{table}[h!]
\centering
\begin{tabular}{ |C{2.5cm}|C{2.5cm}|C{2.5cm}|C{1.8cm}|C{1.8cm}|}
            \hline
            \textbf{Countries} & \textbf{App Name} & \textbf{Tech} & \textbf{Voluntary} & \textbf{Open Source}  \\\hline
            Australia & COVIDSafe & Bluetooth & Yes & Yes  \\\hline
            Austria & Stopp Corona & Bluetooth, Google/Apple & Yes & Yes \\\hline
            Bahrain & BeAware & Bluetooth, Location & No & No \\\hline
            Belgium & Belgium's app & Bluetooth, Google/Apple & Yes & No \\\hline
            Bulgaria & ViruSafe & Location & Yes & Yes \\\hline
            Canada & COVID Alert & Bluetooth, Google/Apple & Yes & Yes \\\hline
            China & Chinese health code system & Location, Data mining & No & No \\\hline
            Cyprus & CovTracer & Location,GPS & Yes & Yes \\\hline
            Czech & eRouska & Bluetooth & Yes & Yes \\\hline
            Denmark & Smittestop & Bluetooth, Google/Apple & Yes & Yes \\\hline
            Estonia & Estonia's App & Bluetooth, DP-3T, Google/Apple & Yes & No \\\hline
            Finland &  Ketju & Bluetooth, DP-3T & Yes & Yes \\\hline
            Germany &  Corona-Warn-App & Bluetooth, Google/Apple & Yes & Yes \\\hline
            Ghana & GH COVID-19 Tracker & Location & Yes & No \\\hline
            
        \end{tabular}
\end{table}
\FloatBarrier

\FloatBarrier
\begin{table}[h!]
\centering
\begin{tabular}{ |C{2.5cm}|C{2.5cm}|C{2.5cm}|C{1.8cm}|C{1.8cm}|}
            \hline
            Gibraltar & Beat Covid Gibraltar & TBD & No & No \\\hline
            Hungary & 	VirusRadar & Bluetooth & Yes & No \\\hline
            Iceland & 	Rakning C-19 & Location & Yes & Yes \\\hline
            India & 	Aarogya Setu & Bluetooth, Location & No & Yes \\\hline
            Indonesia & 	PeduliLindungi	 & TBD & No & No \\\hline
            Iran & 	Mask.ir & Location & Yes & No \\\hline
            Ireland & 	HSE Covid-19 App & Bluetooth, Google/Apple & Yes & No \\\hline
            Israel & 	HaMagen & Location & Yes & Yes \\\hline
            Italy & 	Immuni & Bluetooth, Google/Apple & Yes & Yes \\\hline
            Japan & 	COCOA & Google/Apple & Yes & No\\\hline
            Kuwait & 	Shlonik & Location & No & No \\\hline
            Malaysia & 	MyTrace & Bluetooth, Google/Apple & Yes & No \\\hline
            Mexico & 	CovidRadar & Bluetooth & Yes & No \\\hline
            New Zealand & 	NZ COVID Tracer & QR codes & Yes & No \\\hline
            North Macedonia & 	StopKorona & Bluetooth & Yes & Yes\\\hline
            Northern Ireland & 	Northern Ireland's app & Bluetooth, Google/Apple & No & No\\\hline
            Northern Ireland & 	Northern Ireland's app & Bluetooth, Google/Apple & Yes & No \\\hline
            Norway & 	Smittestopp & Bluetooth, Location & Yes & No \\\hline
            Philippines & 	StaySafe & Bluetooth & Yes & No \\\hline
            Poland & 	ProteGO & Bluetooth & Yes & Yes \\\hline
            
        \end{tabular}
\end{table}
\FloatBarrier

\FloatBarrier
\begin{table}[h!]
\centering
\begin{tabular}{ |C{2.5cm}|C{2.5cm}|C{2.5cm}|C{1.8cm}|C{1.8cm}|}
            \hline
            Qatar & 	Ehteraz & Bluetooth, Location & No & No\\\hline
            Saudi Arabia & 	Tawakkalna & TBD & No & No \\\hline
            Singapore & 	Trace together & Bluetooth, Blue Trace & Yes & Yes \\\hline
            Switzerland & 	Swiss contact-tracing App & Bluetooth, DP-3T, Google/Apple & Yes & No \\\hline
            Thailand & 	Mor Chana & Location, Bluetooth & Yes & No \\\hline
            Tunisia & 		E7mi & Bluetooth & No & No \\\hline
            Turkey & 	Hayat Eve Sığar & Bluetooth, Location & No & No \\\hline
            United Arab Emirates & 	TraceCovid & Bluetooth & No & No \\\hline
            United Kingdom & 	NHS COVID-19 App & Bluetooth, Google/Apple & Yes & Yes \\\hline
            
        \end{tabular}
    \caption{List of countries using different apps and their adopted technologies.}
    \label{tab:countrieslist}
\end{table}
\FloatBarrier

\subsection{Suspended, Replaced and Relaunched Apps}
During the first wave of the pandemic, many countries launched their apps with varying degrees of success. Finland's one million people downloaded Finland's app within just 24 hours after the launch, which is around 20\% of their population. This is mainly because of the high penetration rate of smartphones in the population and prioritized individual privacy using third-party instead of using the government platform \cite{FinlandSucces}. Similarly, Ireland's app was downloaded by 37\% of the population just two months after its launch \cite{FinlandSucces}. This success shows that people value the privacy-preserving attributes of the apps most.

However, apps of some countries are suspended by regulators, some have been replaced with updated versions since their initial launch, and some have been relaunched \cite{SuspendedApps}. Iran's AC19 app was suspended by the Google Play store for allegedly spying on users, and Japan's app was suspended two times due to malfunctioning. Some countries like Norway, Finland, and the UK switched their apps to the Google/Apple framework for its globally available, well-established notification system.

These challenges show that the digital contact tracing apps have several gaps in research for the latest communication technologies, privacy preservation methodologies, etc., which are discussed in the next sections.    
\section{Current Research Gaps} \label{gaps}

Digital contact tracing has the potential of playing a significant role in the current COVID-19 pandemic; however, there are still research gaps to be addressed. Some of the current research gaps are discussed below.  

\subsection{Urgent Need for Well-Designed Prospective Evaluations of Digital Contact Tracing Solutions in Real-World Epidemic Settings}

There is no published or direct evidence to date (as of December 2020) that effectively evaluates different digital contact tracing solutions on offer today in terms of their safety and effectiveness (sensitivity and specificity), interface accessibility, user acceptance, equitable access across different age groups and communities, privacy protection and associated ethical issues when compared with traditional contact tracing in real world epidemic settings \cite{Anglemyer2020}. Despite their promised potential to help identify more contacts, the actual effectiveness of digital contact tracing solutions remains unproven \cite{Braithwaite2020}, and they are unlikely to fit to be used as the sole method of contact tracing to rely upon during an outbreak. Therefore, there is an urgent need to conduct robust research to address these aspects and provide sufficient evidence about how we can best use digital solutions alongside manual methods for optimal epidemic control. 

\subsection{Scalablity of Existing Contact Tracing Systems}

The effectiveness of the current digital contact tracing applications is directly proportional to the adoption rate of these apps by the masses. 
This can be increased through voluntary user adoption by expanding the current IoT infrastructure and providing strong privacy guarantees. Singapore government has recently rolled out their COVID tracking tokens targeting population groups with low smartphone ownership or mastery, especially the elderly who are not tech-savvy \cite{BBC}. Although there are still concerns about user data privacy with the voluntary adoptions of these technologies, scalability can be increased. Users' confidence regarding the preservation of their privacy-sensitive information in current digital contact tracing applications can be strengthened to enhance the voluntary adoption of these technologies. 

\subsection{Improving the technological impediments}
Several technological impediments restrict the use of digital contact tracing technologies. These include inaccuracy in distance estimations using BLE signal strength values, interoperability of different apps based on different architectures, and guarantees regarding privacy and security of data collected through these apps ~\cite{appsurvey}. Research in these areas will complement the digital contact tracing platform's adoption rate to supplement manual contact tracing.   


\section{Emerging Technologies in Digital Contact Tracing} \label{Emerg_Comm}
Several emerging technologies have a high potential of being useful in the context of digital contact tracing. These emerging technologies include Computer Vision (CV), Artificial Intelligence (AI), Machine Intelligence (MI), and various kinds of sensors (ultrasound, visible sensors, and thermal) \cite{IEEE_Access_COVID_Survey_2}. A brief introduction of these technologies and how they can be used for contact-tracing purposes is provided in this section. 

\begin{enumerate}

\item{Computer Vision, Artificial Intelligence and Machine Learning}

Computer Vision (CV) is a technology in which computers are trained to interpret and understand visual images and videos. With the application of AI (e.g., deep learning and pattern recognition), CV can detect and classify any objects using visual data. Using the capabilities of CV based on smart cameras, we can encourage and enforce social distancing by person re-identification for positive cases, estimating distances between persons, and detecting crowds. 
Sometimes the biases in the data can lead to the misrepresentation of the CV algorithms. Therefore, the use of CV application in conjunction with other technologies can provide a reasonable accuracy. 

\item{Ultrasound}

Ultrasonic or Ultrasonic Positioning System (UPS) is an indoor environment communication technology with an accuracy of centimeters \cite{Ultrasound}. In UPS, ultrasonic beacons or nodes are used, which periodically broadcast ultrasonic pulses or Radio Frequency (RF) signals with unique IDs. Using these messages, the users' positions can be detected by the position calculation methods such as trilateration and triangulation \cite{Ultrasonic2}. As compared to RF technologies, these ultrasonic pulses are not affected by electromagnetic interference. However, it is limited to the indoor environment with short-range coverage. Therefore, UPS application in contact tracing can be useful in an indoor environment, especially for high accuracy.   

\item{Visible Sensors}

The emergence of Light Emitting Diodes (LEDs) has provided attractive features of visible lights such as security, privacy, and robustness \cite{VL1,VL2,VL3}. Visible Light Communication (VLC) has two components, which include light transmitters and light receivers. It can provide effective measures in maintaining social distancing while providing precise location and navigation in an indoor environment, especially in the epidemic setting, in monitoring quarantined persons and crowd detection. Due to the low cost and ease of installation VLC receivers can be used as tags and integrated into mobile targets such as shopping carts and robots to avoid the build-up of crowds.  

\item{Thermal}

Thermal-based positioning is classified into two main categories: Infrared Positioning (IRP) and Thermal Imaging Camera (TIC). IRPs systems \cite{TH1,TH2} is considered low-cost, short-range systems (e.g., 10 meters) that can measure the positions of the targets, while TICs can construct the images of the objects from heat emission and operate up to kilometers. By combining IRPs and TICs, social distancing can be monitored in both the indoor and outdoor environments. Based on the images of the TICs, the position between two individuals can be calculated, and crowd forming can be avoided.      

\end{enumerate}

\section{Conclusions} \label{Conclusion}
There has been wide international adoption of digital contact tracing applications in the aftermath of the COVID-19 pandemic. Digital contact tracing is more scalable than manual contact tracing, with the potential of picking contacts that are otherwise untraceable manually, such as encounters with strangers in public transport or a coffee shop. Although several research works have reviewed contract tracing applications and techniques, we present in this paper a thorough analysis of contact tracing applications and techniques in the light of initial deployment experiences of these digital contact tracing technologies and highlight their successes, failures, and pitfalls. Digital contact tracing applications have faced issues, such as low mobile phone penetration, poor user adoption, and privacy concerns, to name a few. In this regard, we discuss how different countries and applications have made different trade-offs and have therefore experienced different amounts of success in effectively combating COVID-19. It is noted that while digital contact tracing apps have their strengths, it is not a panacea; a multi-pronged COVID-19 response requires digital contact tracing along with complements such as manual contact tracing, effective coordination and use of preventive measures such as quarantine isolation, social distancing, hygiene control, and the use of face masks.


\bibliography{main}

\end{document}